# A TEXT-DEPENDENT SPEAKER VERIFICATION APPLICATION FRAMEWORK BASED ON CHINESE NUMERICAL STRING CORPUS

*Litong Zheng, Feng Hong, Weijie Xu*

Shanghai Acoustics Laboratory, Chinese Academy of Sciences, Shanghai
University of Chinese Academy of Sciences, Beijing

## ABSTRACT

Researches indicate that text-dependent speaker verification (TD-SV) often outperforms text-independent verification (TI-SV) in short speech scenarios. However, collecting large-scale fixed text speech data is challenging, and as speech length increases, factors like sentence rhythm and pauses affect TDSV's sensitivity to text sequence. Based on these factors, We propose the hypothesis that strategies such as more fine-grained pooling methods on time scales and decoupled representations of speech speaker embedding and text embedding are more suitable for TD-SV. We have introduced an end-to-end TD-SV system based on a dataset comprising longer Chinese numerical string texts. It contains a text embedding network, a speaker embedding network, and back-end fusion. First, we recorded a dataset consisting of long Chinese numerical text named SHAL, which is publicly available on the Open-SLR website. We addressed the issue of dataset scarcity by augmenting it using Tacotron2 and HiFi-GAN. Next, we introduced a dual representation of speech with text embedding and speaker embedding. In the text embedding network, we employed an enhanced Transformer and introduced a triple loss that includes text classification loss, CTC loss, and decoder loss. For the speaker embedding network, we enhanced a sliding window attentive statistics pooling (SWASP), combined with attentive statistics pooling (ASP) to create a multi-scale pooling method. Finally, we fused text embedding and speaker embedding. Our pooling methods achieved an equal error rate (EER) performance improvement of 49.2% on Hi-Mia and 75.0% on SHAL, respectively.

## 1. INTRODUCTION

Text-dependent speaker verification (TD-SV) requires matching the text content in both enrollment and verification speech, resulting in fixed phonetic information. This leads to better performance metrics like Equal Error Rate (EER) compared to text-independent speaker verification (TI-SV) [1]. TI-SV, being text-agnostic, offers flexibility for various scenarios. However, TD-SV faces challenges like collecting sufficient data for different text requirements, limiting its practical use. In SV, networks like X-vector, ECAPA-TDNN, Conformer, and others [2-6] have shown excellent results in both TD-SV [7-9] and TI-SV [10-12]. Yet, TD-SV has two key challenges: (1) Limited speech data [13] due to varying text requirements, favoring TI-SV. (2) Domain mismatch [1,13], affecting performance due to data and speech disparities.

In order to solve the lack of data, Qin et al. enhanced the TD-SV dataset through text to speech (TTS) with Tacotron2 and voice conversion (VC). [17][18]. To resolve domain mismatch, Qian et al. [12] enhanced system performance by optimizing text-related gradients before pooling operations. They also introduced a speaker-text factorization network [14], which separates speech into a text-independent speaker and text-independent text embedding, then recombines them into a single embedding. Han et al. [15] improved ECAPA-TDNN for short-segment SV with a time-domain multi-resolution encoder. Mak et al. [16] used weight space ensemble for SV challenges, particularly domain mismatch, outperforming traditional methods.

We would like to expand TDSV's identity verification in areas such as financial payments, so the corpus used is a long text random number string such as "8 1 7 3 2 5 9 6 0 4". We segment and sequence the random digital string verification voice in the background to keep it consistent with the order of the registered voice text. In this case, we found that when the text length is long, the model is not very sensitive to the order of the text. an end-to-end TD-SV framework has been introduced. Firstly, data augmentation is carried out through TTS. Secondly, an improved Transformer [19] is utilized to extract text embedding from the augmented dataset. Thirdly, we use the ECAPA-TDNN as the speaker embedding network, and we consider its pooling method ASP overlooks the temporal dimension's importance and is unsuitable for extracting temporal waveforms from fixed-text corpora. We propose a pooling method suitable for TD-SV. Its temporal encoding and compression are accomplished through sliding window pooling in the temporal dimension. Lastly, drawing inspiration from the back-end fusion methods used in the Spoofing Aware Speaker Verification Challenge (SASV2022) [20], text embedding and speaker embedding are either added, multiplied or using CNN fusion [21] to produce the fused result.

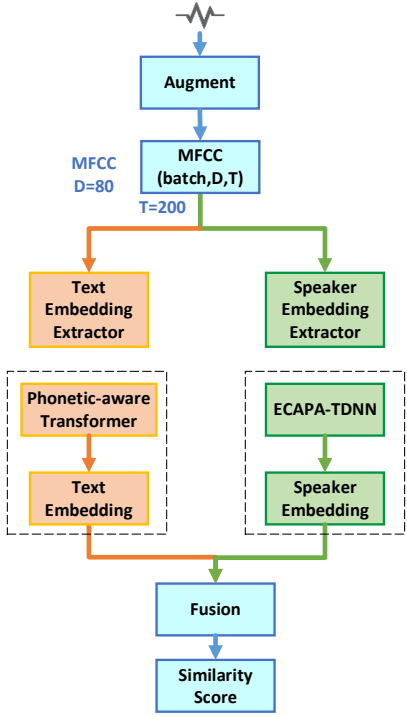

**Fig.1.** *The overall framework diagram of end-to-end TD-SV framework. Above are data augmentation and MFCC feature extraction. On the left and right sides, we have Transformer and ECAPA-TDNN extracting text and speaker embedding, Finally, these embeddings are fused.*

## 2. END-TO-END TD-SV FRAMEWORK

The overall framework diagram of end-to-end TD-SV framework is shown in Fig.1. First, the data undergoes data augmentation, including TTS and speed disturbance. Next, MFCC features are extracted. On the left-hand side is the text embedding extractor, which employs a Transformer with phonetic information. On the right-hand side is the speaker embedding code extractor, utilizing ECAPA-TDNN. Finally, the embeddings obtained from both sides are fused.

### 2.1. Text embedding extraction based on Transformer

In order to better capture the text information in speech, we propose a text embedding network With reference to the work of Qian [14] and speech recognition engine We-Net [22], we employ a single Transformer as depicted in Fig.2. This includes positional encoding, and both the Encoder and Decoder follow standard Transformer structures. The visual representation yields three distinct outputs, each associated with a unique loss function. (1) From the output of the Encoder, we derive a text classifier consisting of a simple CNN block. After passing through CNN, ASP is applied, resulting in a 192-dimensional output serving as text embedding. This is followed by further dimension reduction through a linear layer, with the output dimension matching the number of text labels, assuming a finite set of text labels for classification. Finally, we obtained a loss of $\mathcal{L}_1$. (2) The second part of the output also originates from the Encoder output. It passes through a three-layer CNN to generate input for the CTC loss function. The CTC loss function addresses alignment issues between labels and predictions in neural networks, enabling the model to handle speech with varying pause rhythms. Consequently, it leads to the CTC loss, denoted as $\mathcal{L}_2$. (3) The third part involves Decoder output loss, where phoneme label information is utilized. The phoneme information incorporates pause rhythms during input, and during inference decoding, the results from the text classifier serve as the first decoding hint. This facilitates faster convergence and makes decoded results closely resemble the true text sequence. This yields the loss function $\mathcal{L}_3$. The overall loss size is represented as $\mathcal{L}_{total}$. The specific formula is as follows:

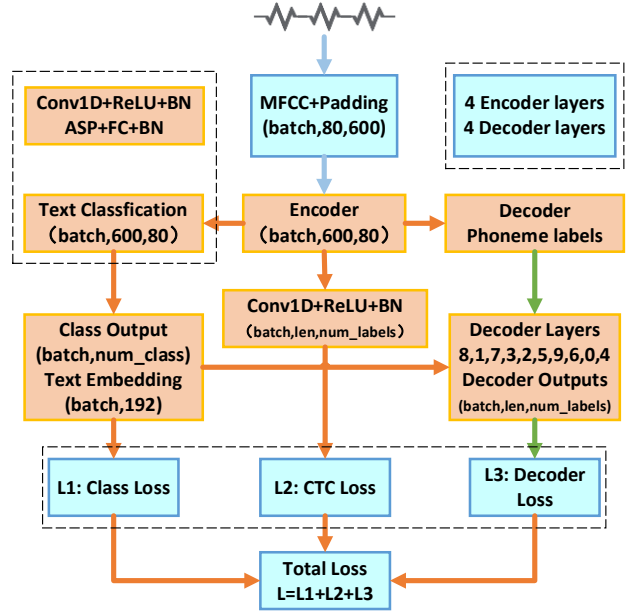

**Fig.2.** *The Text Embedding Extraction Framework Diagram: on the left, we have text embedding extraction and Text Classification Loss; in the middle, there is the CTC Loss; on the right, the Decoder and Decoding Loss.*

$$E_{text} = Pooling(BN(ReLU(Conv1D(X)))) \quad (1)$$
$$\mathcal{L}_1 = CE(BN(FC(E_{text}))) \quad (2)$$
$$\mathcal{L}_2 = CTC(BN(ReLU(Conv1D(X))*3)) \quad (3)$$
$$\mathcal{L}_3 = CE(Y) \quad (4)$$
$$\mathcal{L}_{total} = \alpha\mathcal{L}_1 + \beta\mathcal{L}_2 + \gamma\mathcal{L}_3 \quad (5)$$

Where $X$ is the Encoder output, $E_{text}$ is the text embedding. Pooling in ASP, BN, ReLU and Conv1D combine to a CNN block. In $\mathcal{L}_2$, "*3" represents three CNN blocks. CE stands for Cross-Entropy loss function. In this paper, we emphasize text classification capability, with $\mathcal{L}_{total}$'s three weights being 0.6, 0.2 and 0.2, respectively.

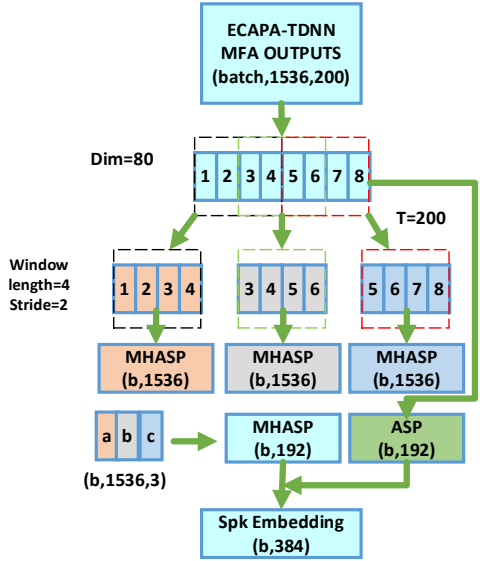

**Fig.3.** *The Sliding Window Attentive Statistics Pooling (SWASP) Diagram: Divide the ECAPA-TDNN MFA outputs into segments based on window length and stride. Apply multi-head attentive statistics pooling (MHASP) to each window segment and concatenate the results. Perform MHASP again on the concatenated results. Finally, combine the obtained results with those from ASP to form the result of multi-scale pooling.*

### 2.2. Speaker embedding extraction and SWASP

The overall computational diagram of SWASP is illustrated in Fig.3. Different colors in the figure represent different sliding windows. If the MFA outputs $\mathcal{M}$ size is [$batch, 1536, 200$], with a window length of 50 and a stride of 25, then $\mathcal{M}$ can be divided along the time dimension into [$\mathcal{M}_{0-50}$, $\mathcal{M}_{25-75}$, $\mathcal{M}_{50-100}$, $\mathcal{M}_{75-125}$, $\mathcal{M}_{125-150}$, $\mathcal{M}_{150-175}$, $\mathcal{M}_{175-200}$]. The size of each $\mathcal{M}_n$ is [batch, 1536, 50]. After applying multi-head attentive statistics pooling (MHASP) to each $\mathcal{M}_n$, we obtain [batch, 1536]. These 7 $\mathcal{M}_n$ segments are fused together, followed by pooling, and the final result is fused with ASP.

For single MHASP, based on the input from the previous layer, three linear transformations are applied to obtain matrices Q, K, and V [19]. The resulting vector after multi-head attention allocation is as follows:

$$Q, K, V = Linear1(X), Linear2(X), Linear3(X) \quad (6)$$

$$A(Q, K, V) = Softmax\left(QK^T / \sqrt{d_k}\right)V \quad (7)$$

$$\alpha = Softmax(\omega_2(tanh(\omega_1 A + b_1) + b_2, dim = 2) \quad (8)$$

$$\mu = \sum_t^T \alpha_t X_t, \; \sigma^2 = \sum_t^T \alpha_t X_t \odot X_t - \mu \odot \mu \quad (9,10)$$

Where $X$ is the pooling input, $A$ is the result obtained after multi-head self-attention, and $\alpha$ represents the weights obtained by ASP for $A$. $\mu$ and $\sigma^2$ are the first-order and second-order statistics obtained from ASP, respectively. $\odot$ is Hadamard Product.

**Table 1.** *SHALCAS22A Recording Information Statistics*

| Text label | Text content | Speech duration |
|---|---|---|
| d001 | 8-1-7-3-2-5-9-6-0-4 | 6s |
| d002 | 8-1-7-3\|2-5-9-6\|-0-4 | 4s |
| d003 | 8-1-7\|-3-2-5\|-9-6-0\|-4 | 4s |
| d004 | 8-1\|-7-3\|-2-5\|-9-6\|-0-4 | 4s |
| d005 | 9-4-0-5\|3-7-2-6\|-8-1 | 4s |
| d006 | 9-4-0\|-5-3-7\|-2-6-8\|-1 | 3s |

### 2.3 Dataset and augment

SHAL dataset includes about 72.3 hours of audio with 46,583 files in 44.1kHz, 16-bit PCM-WAV format. The dataset focuses on speakers aged 10-40 and balances gender. Table 1 summarizes the dataset. We selected 60 individuals, each contributing 25 samples for each text type. Initially, we performed pretraining on 10,000 female voices from AISHELL [24] using Tacotron2 and HiFi-GAN [17]. Then, we applied transfer learning using each individual speaker's speech data, resulting in a total of 60 personalized TTS models. Finally, we conducted TTS with different pause rhythms and introduced speed perturbations of 0.9x and 1.1x, expanding the dataset to approximately six times its original size. Unlike 'Hi-Mia' [24] and 'Ok-Google', the numerical string corpus offers broader applicability.

## 3. EXPERIMENTAL SETUPS

Our experiment is as follows: (1) Performance Evaluation Experiment of Data Augmentation Sets. The network is ECAPA-TDNN, training data is VoxCeleb2 with 5994 speakers. The test sets include VoxCeleb1-O, the original SHAL dataset (Ori), and various combinations of test sets with speed perturbation and TTS enhancement. (2) Experiments involved ECAPA-TDNN with different pooling methods on various evaluation sets, including VoxCeleb1-O, Hi-Mia near-field noise-free dataset, SHAL d002, and d005. (3) SWASP hyperparameters. Network is ECAPA-TDNN, trained and eval on Hi-Mia (4) Text and speaker embedding back-end joint optimization. Except for Experiment (1), the names of the test sets for the other experiments also represent the names of their training sets. We have divided Hi-Mia. SHAL into training and test sets according to 8:2. Each model will first run a pre trained result on Voxceleb2, and then fine tune and test according to their respective tasks.

We drew inspiration from advanced techniques in TI-SV domains, such as inter top-K penalty for optimizing hard samples proposed by Zhao et al [25]. and large margin fine-tuning proposed by T. J [26]. We used Adam optimizer with 0.001 initial learning rate, decayed by 3% every epoch. The loss function was AAM-Softmax, with margin and scale set to 30 and 0.2 [25],In the Text Embedding Extractor, both the Encoder and Decoder have 4 blocks with 4 heads each, and Q, K, V dimensions of 64. In the Speaker Network, the attention mechanism has 2 heads, and only 1 block. All experiments were conducted without noise or reverberation.

**Table 2.** *Experiments on Original Dataset and Various Data Augmentation Combinations: The test network is ECAPA-TDNN, and the training data is Voxceleb2. 'Ori' represents the original SHAL dataset, 'Speed' denotes the augmentation set obtained through data perturbation, and 'TTS' stands for the augmentation set synthesized using Tacotron2.*

| Eval dataset | Sample Nums | EER(%) | MinDCF |
|---|---|---|---|
| VoxCeleb1-O | 37611 | 1.09 | 0.078 |
| Ori | 7140 | 0.69 | 0.050 |
| Ori + Speed | 28800 | 1.44 | 0.118 |
| Ori + TTS | 19200 | 1.93 | 0.245 |
| Speed + TTS | 28800 | 0.27 | 0.025 |
| Ori + Speed+ TTS | 57600 | 1.72 | 0.171 |

**Table 3.** *Different Pooling Strategy Combination Experiments: 'A' represents ASP, 'M' represents single MHASP, and 'S' represents SWASP. 'base' represents baseline.*

| System | VoxCeleb1-O | | Hi-Mia | | SHAL | |
|---|---|---|---|---|---|---|
| | EER(%) | MinDCF | EER(%) | MinDCF | EER(%) | MinDCF |
| A-base | **1.09** | **0.078** | 0.63 | 0.046 | .0.48 | 0.025 |
| M | 1.36 | 0.097 | 1.11 | 0.054 | 0.56 | 0.028 |
| S | 1.35 | 0.095 | 0.79 | 0.070 | 1.67 | 0.100 |
| A+S | 1.18 | 0.083 | **0.32** | **0.033** | **0.12** | **0.011** |
| M+S | 1.29 | 0.088 | 0.84 | 0.071 | 0.35 | 0.056 |
| A+M+S | 1.35 | 0.096 | 0.68 | 0.046 | 0.55 | 0.036 |

**Table 4.** *Hyperparameter Experiments of ASP+SWASP on the HiMia Dataset: 'w' denotes window length, and 's' represents stride.*

| System | EER(%) | MinDCF |
|---|---|---|
| w=25, s=25 | 0.48 | 0.038 |
| w=50, s=50 | 0.36 | 0.041 |
| w=75, s=75 | 0.39 | 0.042 |
| w=100, s=100 | 0.35 | 0.041 |
| w=50, s=10 | 0.33 | 0.041 |
| w=50, s=20 | 0.36 | 0.039 |
| w=50, s=25 | **0.32** | **0.033** |
| w=50, s=30 | **0.32** | 0.040 |
| w=50, s=40 | 0.37 | 0.042 |

**Table 5.** *Fusion of Speaker Embedding from Different Models with the Same Text Embedding. A-VOX2 represents ASP pretrained on VoxCeleb2, while A-SHAL represents pretraining on SHAL. Ultimately, both are used to compute SHAL embedding.*

| System | Addition | Multiplication | CNN fusion |
|---|---|---|---|
| | EER(%) | EER(%) | EER(%) |
| A-Vox2-base | 1.11 | 1.08 | 8.33 |
| A-SHAL | 0.78 | 0.69 | 18.33 |
| S-SHAL | 1.67 | 1.67 | 8.89 |
| (A+S)-SHAL | **0.00** | **0.00** | 10.04 |

## 4. EXPERIMENTAL RESULTS AND ANALYSIS

Table 2. shows ECAPA-TDNN model performance trained on VoxCeleb2 and evaluated with various enhanced sets. It highlights SHAL dataset performance and enhancements. Adding TTS to Ori significantly reduces EER, indicating enhanced speech quality. Enhanced sets themselves display notable EER..

Table 3. ECAPA-TDNN was employed with different pooling methods. Initially pretrained on VoxCeleb2, it was fine-tuned per task and evaluated with task-specific datasets. Indicates that SWASP, designed for text-related tasks, performs worse than the baseline ASP pooling when handling TI training and testing sets, which is expected. If only single MHASP is used, the overall representation capability is weak, leading to a slight performance drop. This issue also occurs with SWASP. The best performance is achieved with the combination of ASP+SWASP by 49.2% on Hi-Mia and 75.0% on SHAL. This combination offers a certain level of TI overall representation capability while maintaining some TD local representation capability. Therefore, it can be concluded from this table that using single MHASP or SWASP may lead to performance degradation due to issues like overfitting. Performance improvement is observed when combined with ASP to form multi-scale pooling.

Table 4. displays hyperparameter experiments for SWASP window length and stride. Pretraining on Voxceleb2 was followed by transfer learning and fine-tuning on Hi-Mia, culminating in Hi-Mia evaluation. The results show that the best performance is achieved with w=50 and s=25. In practice, we observed that an excessively short window length can slow down computation, whereas an overly long one can strain memory resources.

Table 5. presents the back-end fusion of text and speaker embeddings. Three models were selected based on the results in Table 2, using VoxCeleb2 pretraining as the baseline and tested on the SHAL dataset. The test sets also followed the same 8:2 ratio as mentioned earlier. However, the difference here is that the inputs are the extracted embeddings from the network. Two types of text were used, following the SASV2022 approach, three fusion baselines were proposed: embedding addition, embedding multiplication, and embedding CNN fusion. Positive and negative sample pairs were considered not only for speaker categories but also for text categories. As shown in the table, the CNN fusion method exhibited greater variability, with the best-performing embedding multiplication approach. With our own ECAPA+ASP+SWASP as the foundation, the fused EER was 0.00%.

## 5. CONCLUSIONS AND FUTURE WORKS

We've introduced an efficient framework for long numerical text in TD-SV. It covers data augmentation, text and speaker embedding extraction, and dual-embedding code fusion. Our multi-scale pooling method improved performance by 49.2% on Hi-Mia and 75.0% on SHAL. We also introduced a dual-embedding code method, achieving 0.00% EER on SHAL when considering text and speaker labels. Future work focuses on making the method more versatile and optimizing model size for efficiency.